\documentstyle[twocolumn,aps,psfig]{revtex}

\begin{document}

\title{Elusive Unfoldability: Learning a Contact Potential to Fold Crambin}
\author{Michele Vendruscolo and Eytan Domany}
\address{Department of Physics of Complex Systems, 
         Weizmann Institute of Science, Rehovot 76100, Israel}

\address{
\centering{
\medskip\em
{}~\\
\begin{minipage}{13cm}
{}~~~
We investigate the extent to which the commonly used
standard pairwise contact
potential can be used to identify the native fold of a protein.
Ideally one would hope that a {\it universal} energy function exists, 
for which the native folds of {\it all} proteins are the
respective ground states. Here we pose a much more restricted
question: is it possible to find a set of
contact parameters for which the energy of the native contact map of a
{\it single protein} (crambin) is lower than that of  all possible
physically realizable decoy maps. We seek such a set of parameters
by {\it perceptron learning}, a procedure which is guaranteed to find
such a set if it exists.
We found  that it is extremely hard (and most probably, impossible)
to fine tune contact parameters that will assign all alternative conformations
higher energy than that of the native map.
This finding clearly indicates that it is impossible to
derive a general pairwise contact potential that can be used to fold
any given protein. Inclusion of additional energy terms,
such as hydrophobic (solvation), hydrogen bond or multi-body
interactions may help to attain foldability
within specific structural families.
{}~\\
{}~\\
Keywords: protein folding; protein folding potential; 
contact map; neural networks; perceptron.
\end{minipage}
}}

\maketitle

\section{Introduction}

Nearly all the important  biochemical
tasks of organisms such as  catalytic activity, molecular recognition and
transmission of signals are performed by proteins \cite{creighton}.
The biological function of these macromolecules is determined by the
specific shapes into which they fold under physiological conditions.
The blueprint for the protein's conformation is its chemical composition, 
e.g. amino acid sequence. 
The central problem of {\it protein folding } \cite{creighton}
is to predict proteins' native structures from their 
amino acid sequences; 
solution of this problem will 
have a formidable impact on molecular biophysics
and drug design.
At present, genome projects have made available the sequences of
hundreds of thousands of proteins \cite{genome}. 
The full potential of this achievement
will be realized only when we are able to routinely translate the
knowledge of a sequence of a protein into the prediction of 
its shape and function.
Moreover, since by using powerful recombinant DNA techniques \cite{watson}
we can now create  proteins with any pre-designed amino acids sequence, 
it will be possible to create synthetic proteins with entirely novel 
functions.

A conceptually straightforward attempt to solve the problem is to 
construct, for any given molecule, an energy function 
using the inter-atomic potentials  and look for 
energy minima. Alternatively, one can use molecular dynamics, e.g. work 
at an energy corresponding to $kT$ and integrate  Newton's equations.
Such a direct attack on the problem lies beyond the possibilities of
existing computers, partly because of the large number of atoms that
comprise a single protein and partly because the exact potential is not
known (we are looking 
for a classical effective interaction between ions and atoms; furthermore,
folding takes place in the presence of water and the water molecules
must be ``integrated out''). This state of affairs points to a need for 
approximate, coarse grained or reduced 
representations of protein structure and derivation of corresponding 
energy functions.
 
A minimalistic representation of a protein's structure is given by its
{\em contact map} \cite{dill,godzik,holm,md96,vkd97}.
The contact map of a protein with $N$ residues is a $N\times N$ matrix
${\bf S}$, whose elements are defined as
\begin{equation}
S_{ij}=
\left\{
\begin{array}{ll}
1 \qquad  & \mbox{\rm if residues $i$ and $j$
are in contact} \\
0 & {\rm otherwise}
\end{array} \right.
\end{equation}
One can define contact between two residues in different ways.
In this work, we will consider two amino acids in contact when their two
$C_{\alpha}$ atoms are closer than 8.5 \AA \cite{vkd97}.
Given all the inter-residue contacts or even a subset of them,
it is possible to reconstruct quite well a protein's structure, by means of
either distance geometry \cite{ch88}, Molecular Dynamics \cite{brunger}
or Monte Carlo \cite{vkd97}. 

In contrast to Cartesian coordinates, the map representation of protein
structure is independent of the coordinate frame. This property made contact 
maps attractive for protein structure comparisons and for
{\it searching a limited database}  for similar 
structures \cite{dill,godzik,holm}.
A more challenging possibility was proposed recently\cite{md96}: to use
the contact map representation for {\it folding, e.g. to search the 
space of contact maps} for the map that corresponds to the native 
fold. The central premise  of this  program  is that the contact map
representation has
an important computational advantage; 
changing a few contacts in a map induces rather significant large-scale
coherent moves of the corresponding polypeptide chain\cite{vdmoves}.	
The proposed program faces, however, three considerable difficulties:
\begin{enumerate}
\item
One needs an efficient procedure to execute these non-local moves 
\item 
There must be a way to test that the resulting maps correspond to physically
realizable conformations and 
\item
One should construct a reliable (free) energy function, defined in
contact map space, such that low-energy maps can be used to identify the
native one.
\end{enumerate}
We made considerable progress\cite{vdmoves} on the first of these 
problems and have found an efficient method to solve the 
second\cite{vkd97}.	 In this study we present some new questions that
are relevant to the third issue, of identifying a reliable energy
function. We also introduce a suitable methodology to address the
questions raised.

The most commonly used energy function for threading 
sequence  ${\bf a}$
into a fold whose contact map is ${\bf S}$
has the form
\begin{equation}
E({\bf a},{\bf S}, {\bf w}) = 
\sum_{ij} S_{ij} w(a_i,a_j) \; .
\label{eq:contact0}
\end{equation}
The 210 parameters 
$w(a_i,a_j)$ represent the energy gained by bringing amino acids
$a_i$ and $a_j$ in contact.

Of the two main methods that
have been used in the past to derive contact energy parameters,
knowledge-based techniques were the first to be proposed.
These methods rely on the quasi-chemical 
approximation \cite{mj96,hl94,td96,jb96,sjkg97} to derive contact
energies from a statistical analysis of known protein structures.
Although suitable for more limited purposes, 
such as fold recognition \cite{lrw95} or threading \cite{mj96},
energy parameters obtained this way have
failed, so far, to produce acceptable maps  
by  energy minimization (sometimes referred to as {\it
ab initio} folding).

More recently, much attention has been devoted to a new class of potentials,
derived by optimization \cite{mc92,glw92,hs96a,hs96b,ms96,dk96}. 
For example Mirny and Shakhnovich \cite{ms96}
determine the contact energy parameters
by minimizing simultaneously the Z-score of the native maps of a large
set of proteins. Hao and Scheraga \cite{hs96a}, 
using a much more detailed representation,
tried to find energy parameters for which
the native conformation has the lowest energy for a single protein. 

In this work we address the same problem as Hao and Scheraga,
but use the contact representation. That is, we ask whether
{\it it is possible to find contact energy parameters, such that
among all physical maps 
for a particular it single
protein, the energy	of
the native map is the lowest?}
In more detail, one requires that
\begin{equation}
E({\bf a},{\bf S}_0, {\bf w}) < E({\bf a},{\bf S}_{c},{\bf w}).
\label{eq:optimization}
\end{equation}
That is, the parameters	${\bf w}$ should be such that
when the sequence ${\bf a}$ is threaded into  
{\it any} physical non-native contact map ${\bf S}_{c}$, the resulting
energy should be higher than that of the native map  ${\bf S}_0$.

Asking the question posed above in the contact energy representation has
a distinct advantage over other potentials, since in our case the energy
is a {\it linear function} of the parameters ${\bf w}$. Therefore once 
a large library of candidate maps ${\bf S}_c$ has been generated, one can
search, by the well known 
method \cite{rosenblatt,minsky,cover,gardner,nd91} 
of {\it perceptron learning},
for a set of ${\bf w}$ for which Eq.(\ref{eq:optimization}) holds for 
{\it all} maps from this library.

Even though
ideally Eq.(\ref{eq:optimization}) should be satisfied for 
any sequence ${\bf a}$ of amino acids, existing in nature or synthesized,
it is not clear at all that it is possible to 
find a set ${\bf w}$ for which 
(\ref{eq:optimization})	holds for even a single protein.
The reason is that as we
have recently shown\cite{2d},  the number of physically 
realizable contact maps is exponential in the length $N$ of the protein
\footnote{The number of self avoiding walk configurations is
also exponential, albeit with a larger coefficient of $N$ in the 
exponent}.
Thus  for a short protein (with, say,
$N  = 40$),  Eq.(\ref{eq:optimization}) implies that about $2^{40}$ 
conditions should be satisfied by tuning  210 parameters! 

We believe that this is a highly relevant question; clearly the true potential
(which, of course, is far more complex than our Eq (\ref{eq:contact0}))
is able to fold all natural proteins: 
there should be a potential of intermediate complexity between the true
one and the simple contact energy we are testing here, which is able to fold,
say, a family of proteins. This work is a first step towards developing
a methodology to test any such  potential.

\begin{figure}
\centerline{\psfig{figure=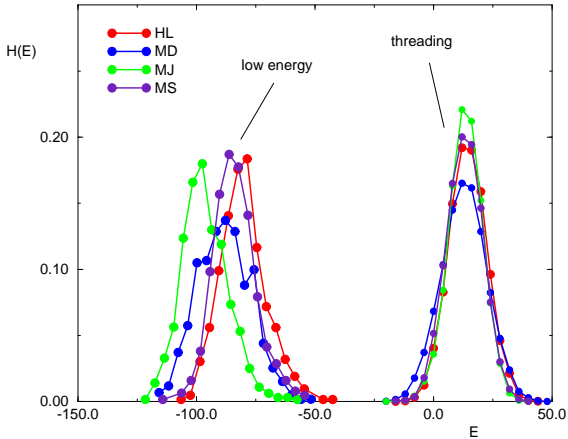,height=7.0cm,angle=270}}
\caption{
Histograms that demonstrate the difference in energy between
ensembles of contact maps obtained by threading and by energy minimization.
We used 4 sets of contact energy parameters:
HL, Hinds and Levitt \protect \cite{hl94}; 
MD, Mirny and Domany \protect \cite{md96}; 
MJ, Miyazawa and Jernigan \protect \cite{mj96};
MS, Mirny and Shakhnovich \protect \cite{ms96}.
}
\label{fig:histo}
\end{figure}

It is important to realize that the conditions we try to satisfy are much
more stringent than the one usually required for successful 
{\it threading} \cite{lrw95,fisher96,finkel97}. 
We did succeed\cite{nvd97} to find a set of contact
energies $\bf w$ that satisfies Eq.(\ref{eq:optimization})
simultaneously for a large family of proteins, 
provided the decoy maps ${\bf S}_c$
were obtained by (gapless) threading. The reason is that
a contact map obtained by threading is, usually, a rather
poor guess for the native fold\cite{lrw95,fisher96,finkel97} 
(see Fig. \ref{fig:histo} of energy histograms of threading and minimization).  

Since one cannot perform an exhaustive search of all physical maps, 
we must {\it generate} a large
ensemble of alternative contact maps of low energy.
We will assume that only such a subset of contact maps
is effectively in competition with the native one
to be the ground state and only these maps
gives rise to
relevant constraints in Eq.(\ref{eq:optimization}).

Our strategy and the outline of this paper are as follows:
\begin{itemize}
\item {\it Generation of alternative conformations:} in
Sec II 
we outline briefly the manner in which such
a  set of low-energy alternatives is generated. 
Details of this method will be presented in a 
separate publication\cite{vdmoves}.
\item {\it Learning of a set of pairwise contact energy parameters:}
in Sec. III, we  present the way in which we use these
contact maps to ``learn'' the energy parameters; the results obtained for a
single protein, crambin, are in Sec IV.
\item
Our results are summarized in Sec V, where we also discuss perspectives
and future directions.
\end{itemize}

We chose crambin as the
particular protein to study since it has a long standing history
in protein folding simulation investigations.
Wilson and Doniach \cite{wd89} used a simplified model in which the
conformation of the backbone and side chains is specified by dihedral angles 
and contact energies are calculated from the distribution of pairwise 
distances observed in known experimental structures.
Among other results, they were able to correctly reproduce the formation 
of secondary structures and many of the features of the hydrophobic core.
Kolinski and Skolnick \cite{ks94} performed accurate Monte Carlo simulations
using a detailed lattice representation, optimized for the prediction
of helical proteins. 
In their model, side chain rotamers were explicitely represented by 
additional single monomers. The energy function, mostly of statistical origin,
contained several terms to help the cooperative assembly of secondary
structures and the packing of the side chains.
On average, their simulation runs ended up in conformations
with the correct topology of the native fold, 
and a RMSD distance of 3 \AA \hspace{3pt}
from the native $C_{\alpha}$ trace .
Hao and Scheraga \cite{hs96a,hs96b} showed, 
by optimizing an extended set of energy parameters, that
it is possible to fold crambin
within 1-2 \AA \hspace{3pt} RMSD from the native state.	
Their conclusion is, however, 
that it is always possible to find structures with lower
energy than the native state.

It should be borne in mind that
within the contact map representation, conformational
fluctuations, as measured by RMSD, amount to 1.1 \AA \hspace{3pt} 
for crambin. This result is obtained by constructing 1000 structures, 
following the method described in Ref. \cite{vkd97}, 
all with contact maps {\em identical}
to the native one, and averaging their RMSD values.

Within the contact energy model, 
existing sets of contact potentials  
perform very poorly in a computer experiment of folding crambin.
We demonstrated this by performing 
a Monte Carlo minimization, using these potentials, 
starting from the experimentally known native structure.
For all the contact potentials tested this procedure identifies
easily conformations that are very different from the native one,
and of significantly lower  energy (see Fig. \ref{fig:histo}).
This clearly proves that for these potentials energy minimization
will lead to non-native states and also demonstrates that
our minimization procedure yields maps of much lower energy than those obtained 
by threading.

\section{Generation of alternative conformations}

\begin{figure}
\centerline{\psfig{figure=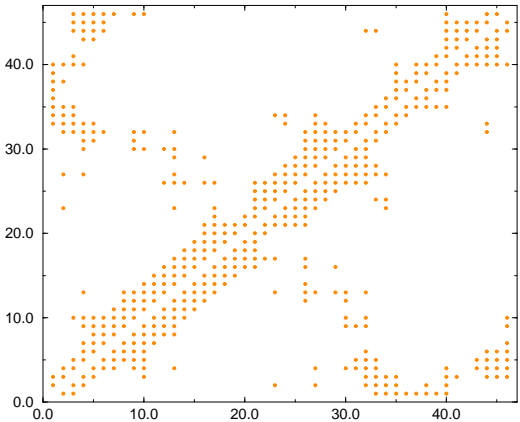,height=7.0cm,angle=270}}
\caption{Contact map for the native state of  crambin. Dark dots represent 
contacts ($S_{ij}=1$).
There are 187 non-nearest neighbors contacts.}
\label{fig:crambin-map}
\end{figure}

Crambin \cite{crambin} is a protein of length $N=46$; its native map,
constructed by taking the coordinates of the 
$C_{\alpha}$ atoms from the PDB and using a threshold
of 8.5 \AA \hspace{3pt} to define contacts, 
is shown in Fig.\ref{fig:crambin-map}. 
Our aim is to generate 
maps of energy low enough  to ``compete'' with the native contact map. 
These alternative candidate maps should be markedly different from one another, 
e.g. have a large relative Hamming distance \cite{vkd97}
\begin{equation}
D^{{\rm map}} = \sum_{j>i} | S_{ij} - S_{ij}^{\prime} | \;.
\label{eq:hamming}
\end{equation}
The number of contact maps that can be actually sampled 
in any reasonable time is a negligible fraction of the
$O(e^{46})$ physical maps.
Therefore, moving in an efficient way in the space of physical contact maps is
an essential component of our program. Clearly, by turning existing
contacts off and non-existing ones on, we can generate large scale
moves of the polypeptide chain; moves that would have taken very
long time to accomplish in real space. This is one of the most attractive
features  of working with contact maps. There are two main problems
with doing this. First, if one selects at random the contacts that are 
to be modified, chances are that the resulting maps 
will not be physical\cite{md96}: that is, there
exists no real polypeptide chain conformation whose contact map is the
one we found. The second 
problem is that we would like to work with moves that do
not destroy secondary structure elements ($\alpha$-helices and $\beta$-sheets).

As can be seen from Fig.\ref{fig:crambin-map}, these appear as
clusters of non-zero entries in the map. 
Thick bands along the principal diagonal represent $\alpha$ helices;
the small antiparallel $\beta$ sheet which characterizes the native fold
of crambin appears as a thick band, perpendicular to the principal diagonal, 
whose two strands extend from amino acids 1 to 4 and from 32 to 35, 
respectively.
A contact map is roughly characterized by the number and the
respective positions of its secondary structure elements.
A typical native map has, in addition,
isolated  entries (single contacts or small clusters) that contain information
about the global fold and relative positions 
of the secondary structure elements.

We  present here only a short description of our Monte Carlo method; 
for a more detailed exposition we refer the reader to Ref. \cite{vdmoves}.
Our algorithm consists of three steps.
The first step 
consists of {\it non-local moves}. We start by  identifying ``clusters''
of contacts in an existing map. These clusters represent
either $\alpha$-helices or $\beta$-sheets (parallel or anti parallel), or
small groups of contacts between amino acids that are well-separated along
the chain. The clusters 
are identified on a given map by laying down bonds 
that connect neighboring contacts
on the map and identifying clusters of contacts that are 
connected by such bonds \cite{vdmoves}.
Some of the existing clusters of contacts are removed and some other groups are
restored elsewhere. 
This way secondary structure elements are destroyed and recreated at
different locations and orientations.
The ``energy'' of the resulting coarse map is evaluated and
a low energy map is retained. This map serves as 
the starting point for the second step:
{\it local moves}. This is a refinement procedure 
that consists of turning on or off
(mostly one at a time) contacts that 
are in the vicinity of existing ones, following some of the rules
introduced in Ref. \cite{md96}.
Again, only moves that lower (or do not significantly raise) 
the energy are accepted.
These first two steps are fast operations, since 
they involve binary variables.
Most of the computer time is taken by the third step, {\it reconstruction},
where we deal with the major problem 
of ensuring that we stay in the subspace of physical maps.

A generic contact map is not guaranteed to correspond to any real conformation
of a polypeptide chain in space. To solve this problem, we developed
an efficient Monte Carlo reconstruction method that checks 
whether any given target map is physical or not\cite{vkd97}. This is done 
by working with a string of beads that represents
the backbone of the polypeptide chain. The beads are moved around without 
tearing the chain and without allowing one bead to 
invade the space of another. The motion of this string
is controlled by a ``cost function'' which vanishes 
when the contact map of the string
coincides with that of the target map. The cost 
increases when the difference between
the two maps increases. This procedure ends up with 
a chain configuration whose
contact map is physical by definition and  close 
to the target map. Thus we are
able to efficiently ``project'' any map that we have generated 
in the first two steps
onto the subspace of physical maps.

Having described the manner in which a single low energy chain and its corresponding 
map are obtained, we
turn to describe the manner in which we generated
a representative ensemble of contact maps, to be used in the derivation
of contact energy parameters.
In general, one expects to have two interplaying levels of optimization.
On the one hand, one has to satisfy Eq.(\ref{eq:optimization}) for
contact maps that are very different from the native one.
On the other hand, with the same set of energy parameters, one should
be able to discriminate between the native contact map and those maps that are
close to it.          
We generated conformations close to the native one by running a series
of Monte Carlo minimizations, {\it starting from the native state}. 
This procedure
was {\it not} carried out in contact map space, but rather by using
a local Monte Carlo procedure on the backbone or chain 
of beads described above, with the position of each bead 
defined in real space; the 
elementary move of this procedure is of the 
crankshaft type \cite{ssk94}.
Each minimization consist of $N_{LMC}$ such Monte Carlo steps
and yields a chain and its low-energy candidate contact map.
A move is accepted according to the Metropolis prescription at a given
fictitious temperature $T_{LMC}$.
The procedure is then repeated, starting again
from the native state but using different random numbers and generating a 
different map.
We call this procedure $D_1$; it generates an ensemble of low energy maps 
that are (relatively) close to the native fold.

\begin{figure}
\centerline{\psfig{figure=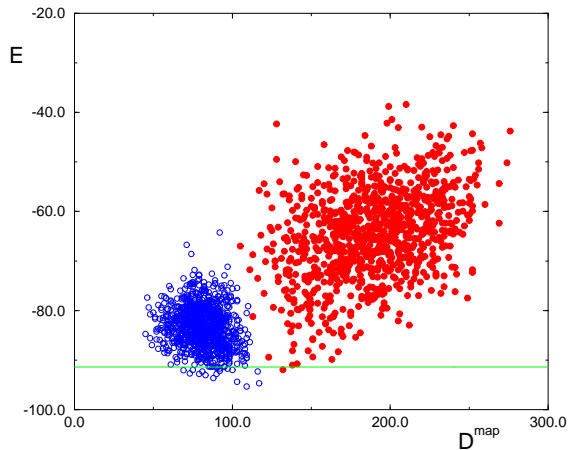,height=7.0cm,angle=270}}
\caption{
Energy $E$ versus Hamming distance $D^{\rm map}$ for 1000 contact maps
obtained by procedures $D_1$ and $D_2$. 
The contact maps are generated using the energy parameters of epoch
$t=9$ (see next section) and $N_{LMC}=10^6$.
The line corresponds to the energy of the native contact map of crambin.
}
\label{fig:ed}
\end{figure}

To generate conformations far from the native one, 
we use the three-step Monte Carlo method described above, 
supplemented by a fourth step;
 a further real space Monte Carlo minimization, as in procedure $D_1$.
That is, the three-step algorithm that performs global moves in contact map
space ends with a chain conformation; this is used as the initial state of 
the $D_1$ procedure (again using $N_{LMC}$ local steps).
Each search gives rise to a low energy contact map.
Such a contact map is used as the staring point for the following global
search. This second procedure, called $D_2$, generates a set of low energy
maps that are very different from the native fold.
In Fig. \ref{fig:ed} we show
energy versus Hamming distance of typical contact maps 
obtained by procedures $D_1$ and $D_2$.

The energies and maps obtained
by the two minimization procedures depend strongly on the length of the run.
As a part of our strategy, we repeated procedures $D_1$ and $D_2$
for $N_{LMC}=10^3,10^4,10^5,10^6,10^7$ local steps.
By combining the two sets of contact maps derived by the two 
minimization procedures, we obtained a representative set of $P$ contact maps.

\section{Learning contact energies}
\subsection{Derivation of contact energies as a learning problem}
In the crambin\cite{crambin} chain, 
5 out of the 20 amino acids do not appear and 3 appear only once. 
Thus, among the corresponding 210 possible contact energies,
only 117 parameters can effectively enter the energy (\ref{eq:contact0})
for any set of candidate maps.
These parameters form a 117-component vector $\bf w$.
The native map of Fig.\ref{fig:crambin-map} contains 187 non-nearest neighbor
contacts, 
which involve only 72 of the 117 possible contact energy parameters.
We now show that for any map ${\bf S}_\mu$ the condition 
Eq.(\ref{eq:optimization})
can be trivially expressed as 
\begin{equation}
{\bf w} \cdot {\bf x}_\mu > 0
\label{eq:newopt}
\end{equation}
To see this just note that for any map ${\bf S}_\mu$ the energy  
(\ref{eq:contact0})
is a linear function of the 117 contact energies that can appear and it
can be written as 
\begin{equation}
E({\bf a},{\bf S}_\mu, {\bf w}) = 
\sum_{c=1}^{117} N_c({\bf S}_\mu) w_c
\label{eq:newener}
\end{equation}
Here the index $c=1,2,...117$  labels the different contacts that can appear
and $N_c({\bf S}_\mu)$ is the total number of contacts of type $c$ that actually
appear in map ${\bf S}_\mu$. The difference between the energy of this map and
the native one is therefore 
\begin{equation}
\Delta E_\mu = 
\sum_{c=1}^{117} x^\mu_c w_c = {\bf w} \cdot {\bf x}_\mu 
\label{eq:Ediff}
\end{equation}
where we used the notation
\begin{equation}
 x^\mu_c = N_c({\bf S}_\mu)- N_c({\bf S}_0)
\label{eq:Ndiff}
\end{equation}
and ${\bf S}_0$ is the native map.  

Each candidate map ${\bf S}_\mu$ is represented by a vector ${\bf x}_\mu$ and
hence the question raised in the introduction becomes 
whether one can find a vector
$\bf w$ such that condition (\ref{eq:newopt}) holds for all ${\bf x}_\mu$? If such
a $\bf w$ exists, it can be found by {\it perceptron learning}.

\subsection{Perceptron: Learning from examples}

A perceptron is the simplest neural network \cite{rosenblatt}.
It is aimed to solve the following task. Given $P$ patterns 
(also called input vectors, examples) ${\bf x}_{\mu}$,
find a vector ${\bf w}$ of weights, 
such that the condition
\begin{equation}
h_{\mu} = {\bf w} \cdot {\bf x}_{\mu} > 0
\label{eq:perceptron}
\end{equation}
is satisfied for every example from a training set of $P$ patterns,
${\bf x}_{\mu}$, $\mu=1,\ldots,P$. 
If such a $\bf w$ exists for the training set, the problem is {\it learnable};
if not, it is unlearnable.
We assume that the vector of ``weights''
$\bf w$, as well as all examples ${\bf x}_{\mu}$ are normalized,
\begin{equation}
{\bf w} \cdot {\bf w} = {\bf x}_{\mu} \cdot {\bf x}_{\mu} = 1
\end{equation}
The vector ${\bf w}$ is ``learned'' in the course of  a training session.
The $P$ patterns are presented cyclically; after presentation of pattern $\mu$
the weights ${\bf w}$ are updated according to the following learning rule:
\begin{equation}
{\bf w}^{\prime} = \left\{ 
\begin{array}{ll}
\frac{{\bf w} + \eta {\bf x}_{\mu}}
                       {|{\bf w} + \eta {\bf x}_{\mu}|} \qquad & 
{\rm if} \qquad  {\bf w} \cdot {\bf x}_{\mu} <0 \\
 & \\
~~ {\bf w} & {\rm otherwise}
\end{array}
\right.
\label{eq:learnrule}
\end{equation}
This procedure is called learning since when the present $\bf w$ misses
the correct ``answer'' $h_\mu >0$ for example $\mu$, all weights are 
modified in a manner that reduces the error. No matter what
initial guess for the $\bf w$ one takes,
a convergence theorem guarantees that if a solution ${\bf w}$ exists,
it will be found in a finite number of training steps.
\cite{rosenblatt,minsky}.

Different choices are possible for the parameter $\eta$.
Here we use the learning rule introduced in Ref.\cite{nd91},
since it allows, at least in principle, to assess whether the problem
is learnable.
The parameter $\eta$ is given at each learning step by
\begin{equation}
\eta = \frac { -h_{\mu} + 1/d } { 1-h_{\mu}/d }
\end{equation}
where the parameter $d$ (called despair) evolves during learning
according to
\begin{equation}
d^{\prime} = \frac { d + \eta } { \sqrt{1 + 2 \eta h_{\mu} + \eta ^2} } \; .
\end{equation}
Initially one sets $d=1$.

The training session can terminate with only two possible outcomes.
Either a solution is found (that is,
no pattern that violates condition (\ref{eq:perceptron}) is found in a cycle),
or unlearnability is detected. The problem is {\it unlearnable} if
the despair parameter $d$ exceeds a
critical threshold\cite{nd91}
\begin{equation}
d > d_c = \frac { M^{M+1} } { 2^{M-1} } \; ,
\label{eq:dc}
\end{equation}
where $M$ is the number of components of ${\bf w}$.

Evidently, once the requirement posed in the introduction, 
Eq (\ref{eq:optimization}), has been expressed in the form (\ref{eq:newopt}),
the question whether it does or does not have a solution reduces to deciding
whether a set of examples is learnable by a perceptron.
Every candidate contact map (generated by the search procedure
described above) provides a  
pattern for the training session. Note that the vector ${\bf x}$ defined
in Eq. (\ref{eq:Ndiff}) must be normalized before 
(\ref{eq:learnrule}) is used. Before turning to present our results for
the learnability of crambin using $M=117$ contact parameters, we address the
same question but use a much simpler potential, that of the HP model\cite{HP}

\subsection{Can the HP model stabilize crambin?}
We tried to stabilize the native map of crambin using the parametrization
of the HP model. This has an even simpler potential than the one we use;
whereas
we have 117 contact energies to tune, the HP model attempts to
satisfy Eq. (\ref{eq:optimization}) using only $M=3$ parameters, $w_1,w_2,w_3$. 
The perceptron learning procedure detected clearly and unambiguously that
this is an unlearnable problem.

\begin{figure}
\centerline{\psfig{figure=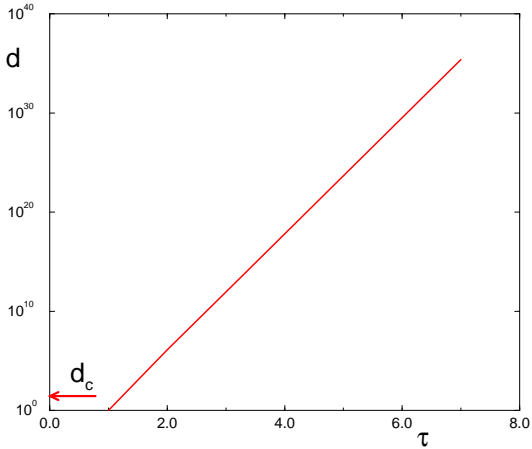,height=7.0cm,angle=270}}
\caption{
Increase of the despair $d$ with the number $\tau$ of learning sweeps
for the HP learning.
}
\label{fig:hp}
\end{figure}

We relabeled the amino acids in the crambin sequence 
following the usual partition  into hydrophobic (H) and polar (P) residues.
Examples were generated
and then 
the perceptron learning procedure was applied 
to this  problem. We measured the value of the despair $d$ as learning
progressed; the result is shown in 
Fig. \ref{fig:hp}. The critical value $d_c=3^4/2^2$
was reached rather fast when we used only 306 examples;
that is, we established that the problems is unlearnable. 

This is a very important lesson from a simple model: 
with only 2 species of amino acids, the contact map
representation is not suitable 
for folding crambin since it is {\em impossible}
to find a set of contact energy parameters that 
satisfy Eq.(\ref{eq:optimization}).
We regard the present work, which employs contact energy parameters
between the 20 existing species of amino acids (15 for crambin), 
as the first step
in a progressive and controlled improvement of protein structure representation.

The question we posed concerns learnability of 
an exponential number of examples, of which we can sample only a small subset.
When 15 amino acids are used the problem is far from being as simple as
the HP model; this is evident from the fact that a large number of examples,
generated by threading, can be easily learned. 
Hence, in order to answer the 
question we posed
it is important to choose a strategy that generates ``hard'' examples. Such
a strategy
is described next. 

\subsection{Iterative Learning and Generation of Examples}

The contact energy parameters are learned 
in an iterative manner, i.e. examples are generated and then learned; the new
$\bf w$ is then used to generate new examples and so on. 
We will refer to each such iteration  as an {\em epoch}.
At epoch $t=0$ we start from an arbitrary set ${\bf w}_0$ of parameters;
we used those that were derived in Ref. \cite{vkd97}
for the present $C_{\alpha}$ representation.
Using the procedure of generation of low energy conformations
discussed above with these energy parameters,
we generated a set of $P_0$  low energy contact maps. This completes epoch 0 and
we can  start epoch $t=1$.
The $P_0$ low energy contact maps obtained in epoch 0
constitute the training set  to learn
new contact parameters, ${\bf w}_1$, according to the perceptron
learning rule discussed in the previous section.
Using the newly derived energy parameters, we generate a new set of $P_1$
low energy contact maps. This set is added to the old training set so that
now we have $P_0+P_1$ examples, all of which are learned in 
epoch $t=2$. In the present work
this iterative procedure is repeated up to epoch 10. 

\section{Elusive Unlearnability}
\subsection{Impracticality of despair} 
\begin{figure}
\centerline{\psfig{figure=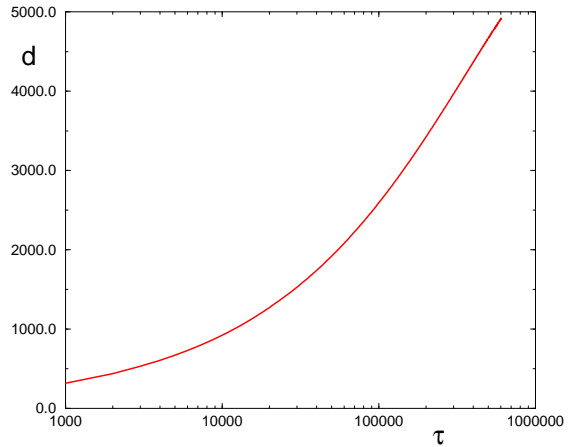,height=7.0cm,angle=270}}
\caption{
Increase of the despair $d$ with the number $\tau$ of learning sweeps
for a typical case of learning dealt with in this work.
}
\label{fig:despair}
\end{figure}

We summarize here our main result about the  question we have addressed 
in the present work. We will present below considerable
evidence supporting our main conclusion: 
\begin{quote}
the problem of fine tuning  the contact energy parameters to stabilize 
the native state of crambin is {\em effectively unsolvable}.
\end{quote}
By this we mean that the problem is either unlearnable, or learnable with 
a learning time which exceeds any realistic scale. We cannot give a clear-cut
answer as we did for the HP model since the 
condition that should be met\cite{nd91} to establish 
unlearnability is numerically impractical; according to 
Eq.(\ref{eq:dc}), for $M=117$, the critical despair $d_c \simeq 10^{87}$.
Moreover,
from Fig.\ref{fig:despair} we see that in this particular problem,
when we tried to learn $P=195124$ examples, the
despair $d$ increases {\em logarithmically} with the learning time $\tau$.
This particular learning task took $\tau = 606756$ sweeps to be solved,
and the final value of the despair was $d=4921.01$.
The size of $P$ and $\tau$$P$ and $\tau$  is strikingly larger than those
involved in the HP case, where $\tau = 2 $ sweeps and $P=306$ examples
were enough to obtain an unambiguously negative answer, (in that case
after $\tau = 7$ the despair was $d > 10^{30})$.

This is also in contrast to perceptron learning of 
$P>2M$ randomly generated examples (which is an unlearnable 
problem\cite{cover,gardner} for large $M$); there
$d$ grows {\em exponentially} with
learning time \cite{nd91}.
Since the time that would be needed to exceed the critical despair
in our particular problem is beyond any reasonable estimate, we have to 
resort to alternative non-rigorous  ways to test learnability of this task.

\subsection{Generalization error}

As explained above, the parameters $\bf w$ that were 
obtained after a new epoch solve a larger training set of 
examples and, hence, they may well be a ``better solution'' of the problem.
The quality of any solution $\bf w$  is measured by
the {\it generalization error} $\varepsilon_g$. To determine it,
one generates a set of new
examples that were not used 
in the training procedure; $\varepsilon_g$ is the fraction of 
examples for which $\bf w$ produces the 
wrong answer and it should decrease when the
training set is increased.

\begin{figure}
\centerline{\psfig{figure=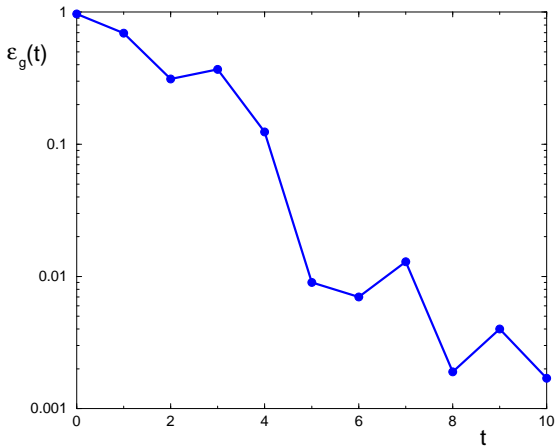,height=7.0cm,angle=270}}
\caption{Generalization error
$\varepsilon_g(t)$ of conformations with energy lower that 
the native state as a function of the epoch $t$. 
}
\label{fig:qlearn}
\end{figure}

In the context of our problem we generate at epoch $t$,
using the procedure described above with the current 
energy parameters ${\bf w}_t$,
a set of $P_t$ of low-energy contact maps. $\varepsilon_g(t)$ is 
the fraction of those contact maps 
that violate Eq.(\ref{eq:optimization}) 
and hence have lower energy than the native map.
The dependence  of $\varepsilon_g(t)$ 
on the epoch index $t$ is shown in Fig. \ref{fig:qlearn}.

Initially $\varepsilon_g(t)$ decreases drastically with  $t$.
We used several of the existing knowledge-based contact potentials 
\cite{md96,mj96,hl94,ms96} 
as our starting energy parameters ${\bf w}_0$; 
the fact that $\varepsilon_g(0) \simeq 1$ signals that these potentials
fail completely the test of assigning the native map an energy that 
is lower than that of maps obtained by our
minimization procedure. This is to be compared with the good performance of
the same potentials on testing the native fold against maps obtained by	 
{\it threading} \cite{nvd97}, highlighting the point made in the Introduction,
that stabilizing the native map against 
our low-energy decoys is a much 
more difficult challenge than  stabilizing it against
maps obtained by threading.

With increasing epoch index, however, the generalization error
does not level out at zero; rather, it fluctuates
at the level of 0.2 - 1 percent. Complete learning is elusive;
this behavior indicates that the problem is, probably, unlearnable.

\subsection{Learning time}
\begin{figure}
\centerline{\psfig{figure=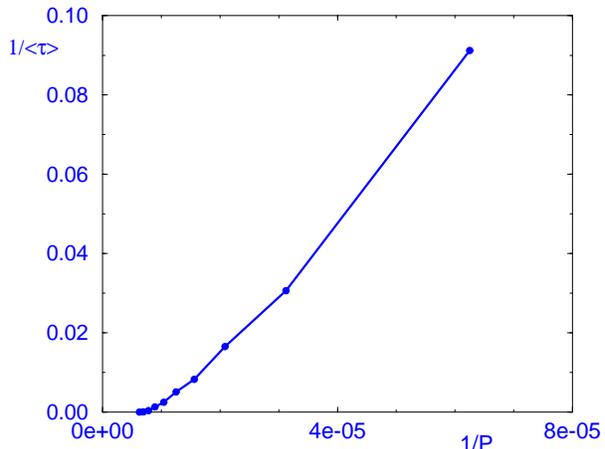,height=7.0cm,angle=270}}
\caption{Inverse average learning time $1/\langle \tau \rangle$ 
as a function of the inverse number $1/P$
of examples.
}
\label{fig:tau}
\end{figure}

Further evidence supporting unlearnability
comes from the way the
learning time $\tau$ increases with the size of the training set.
$\tau$ is defined as the number of sweeps through
the entire training set that is needed to learn all the $P$ examples.
In an unlearnable case there exist  sets of examples for which no solution
can be found; 
for large enough $P$  the training set will include, with non-vanishing
probability, such an unlearnable
subset. This means that the learning time
$\tau$ diverges for a finite $P$.
We show in Fig. \ref{fig:tau} the average inverse learning 
time $1/\langle \tau \rangle$ 
as a function of the inverse number $1/P$ of examples. 
The curve was  obtained as follows.
At the end of our last epoch we collect
all $P^{\rm tot}$
contact maps that have been accumulated so far (during all epochs).
Of these we randomly select a subset of $\Delta P$ maps
and  compute the learning time $\tau$.
This process is repeated $N_L$ times, each time selecting a different set
of $\Delta P$ examples.
$\langle \tau \rangle$
is  the average learning time of these $N_L$ learning processes.
To study how the learning time $\tau$ increases with the 
number of training examples, we repeat the previous procedure,  
randomly selecting $N_L$ training sets of $P=n \Delta P$  patterns 
in each and compute the average learning time as before.
In the data shown in Fig. \ref{fig:tau} we used $N_L=6$ at epoch $t=9$.
We followed this random selection procedure to eliminate all possible dependence
of the learning time on the epoch index,  isolating the variation of $\tau$ with
the size of the training set.

The observed increase of the learning time  with $P$ is 
compatible with a divergence of $\langle \tau \rangle$ at some finite 
$P_c \approx 5\cdot 10^5$, again indicating unlearnability.

\section{Analysis of the contact energies}
Even though the problem is apparently unlearnable, our procedure
produces contact energies that have several appealing features.
One of these has been mentioned above: whereas for the existing contact 
potentials it is very easy to find maps whose energy is below that of
the native map, with the $\bf w$ obtained after several training epochs 
this becomes a difficult (albeit possible) task (see Fig. \ref{fig:qlearn}). 
That is, the generalization error becomes very small. We present now some 
other features of the contact parameters obtained by our learning procedure.

\subsection{Energies of the false ground states}

Another measure of the quality of the energy parameters is given by the
gap $\Delta E$ 
between the energies of the false ground states and  that of the 
native  map. A negative value of $\Delta E$ means that 
(\ref{eq:optimization}) is violated.

\begin{figure}
\centerline{\psfig{figure=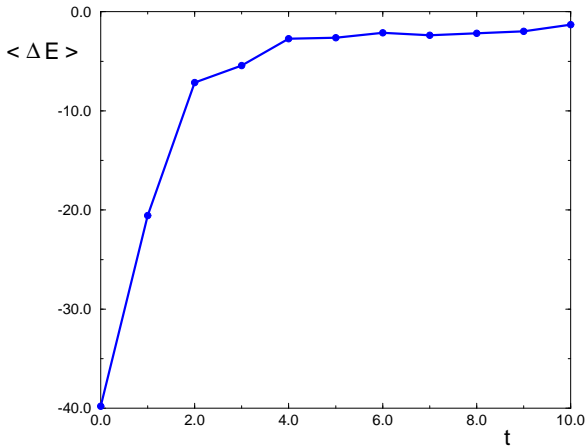,height=7.0cm,angle=270}}
\caption{Energy gap $\langle \Delta E \rangle$
of the violating examples at successive epochs.}
\label{fig:alfv}
\end{figure}

We found that the average $ \vert \Delta E \vert$
of the violating examples decreases with the epoch index, 
see Fig. \ref{fig:alfv}. 
Hence our learning procedure flattens the energy landscape,
reducing both the number of violating examples and their gap.
Another relevant question concerns the ''location'' of these false minima,
i.e. how different are the corresponding structures from the native one? 
To study this, we reconstructed the three dimensional
conformations corresponding to the violating examples and 
measured their average RMSD distance from the native conformation. 
We found that the RMSD does not decrease with the epoch index;
moreover, using procedure $D_2$, false minima are 
found at an approximate average RMSD of 10 \AA \hspace{3pt} 
at {\em any} epoch.
Only their number decreased significantly.
Hao and Scheraga \cite{hs96a}, on the other hand, reported that the distance 
of their false ground states from the
native conformation did decrease as their 
energy parameters became better optimized. 

\subsection{Energy distribution at successive epochs.}

\begin{figure}
\centerline{\psfig{figure=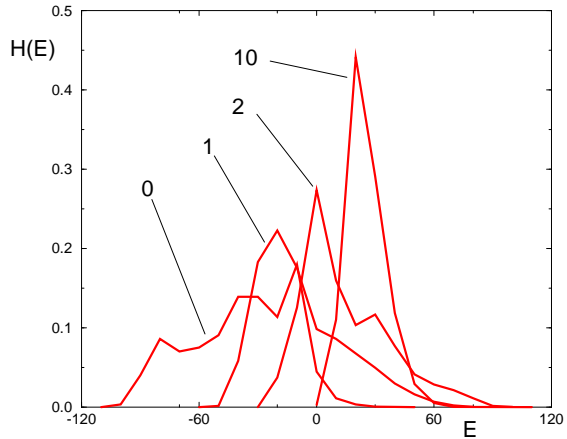,height=7.0cm,angle=270}}
\caption{
Normalized histogram $H(E)$ of the energies of the contact maps
at epochs $t=0,1,2$ and 10. The energy scale is shifted so that
the native contact map energy is 0.
}
\label{fig:he}
\end{figure}

As already observed, 
with the initial energy parameters the vast majority of the contact maps
that are generated have an energy lower than the native contact map.
As can be seen from Fig. \ref{fig:he}, where the energy scale is shifted
so that the native contact map has always zero energy,
for increasing epoch index, the energy distribution shifts to the right
and becomes narrower. Learning is thus accompanied by an improvement
of the $Z$-score, which is a commonly used estimator of the quality of a set
of energy parameters \cite{ms96}.

\subsection{Allowed region in parameter space}

The vector ${\bf w}$ of energy parameters lies on the surface of a unit
sphere in the $N_{{\bf w}}=117$ dimensional parameter space. 
Each example introduces a hyperplane 
which restricts the allowed vectors to one of its sides. The region which
satisfies all the $P$ constraints is called {\it version space}. All vectors
$\bf w$ that lie within version space are compatible with the 
constraints given by 
Eq. (\ref{eq:optimization})
and, therefore, are
solutions of the learning problem.
As more
examples are added, version space may shrink - if the problem is unlearnable,
the (relative) volume of version space shrinks to zero.

\begin{figure}
\centerline{\psfig{figure=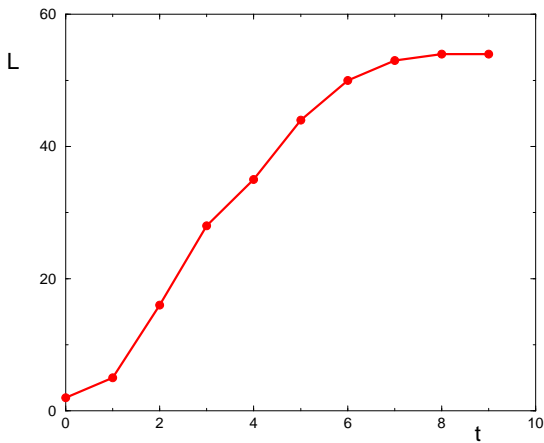,height=7.0cm,angle=270}}
\caption{Number $L$ of locked directions in parameter space as a function
of the epoch $t$.
}
\label{fig:locked}
\end{figure}

To estimate the size of version space  
we generated an ensemble of solutions
by the following Monte Carlo sampling.
At each epoch we arrive by perceptron learning 
at  a particular solution $\bf w$ that satisfies the 
set of $P$ examples which define the current version space.
Starting from this solution, we performed
a random walk on the surface of the unit sphere of $\bf w$ vectors.
The elementary step is to choose at random a component $k$ of ${\bf w}$ and
to change it,
\[
w_k^{\prime} = w_k + \varepsilon
\] 
where  $\varepsilon$ is
a random displacement.
The new array of weights is kept (and normalized) 
if it is still a solution, otherwise
it is rejected. This updating rule is repeated many times, and
eventually a sizeable number of different solutions is obtained.
Next we
perform a principal component analysis of the covariance matrix 
of this ensemble of solutions.
The covariance matrix is defined as
\begin{equation}
C_{ij}= \langle ( w_i - \langle w_i \rangle )
                ( w_j - \langle w_j \rangle ) \rangle  \; ,
\end{equation}
where $\langle  \cdot \rangle$ denotes averages  taken 
over the ensemble of solutions.
Let $\lambda_i >0$ and ${\bf v}_i$ be the eigenvalues and the corresponding
eigenvectors of the covariance matrix.
Clearly, $\sigma_i = \sqrt \lambda_i$ is the standard deviation
which measures the spread of our ensemble of solutions ${\bf w}$  along 
direction ${\bf v}_i$.
If we observe $\lambda_i \rightarrow 0$, this means that along the 
corresponding direction the width of version space has shrunk to zero.
The projections of our vector of energy parameters along these directions 
are fixed and cannot
be changed. 
As shown in Fig. \ref{fig:locked}, the number $L$
of directions whose corresponding
eigenvalues approach zero increases with the epoch index $t$ (to about half
the number of directions).

We also checked whether the directions
that are constrained do or do not change with the epoch index 
and found that these directions
become conserved. The check was performed by measuring for $t > 6$
the variance of the parameters along the directions that were 
locked at epoch $t=6$.
Hence further optimization of these parameters to fold other proteins
is ruled out.

\begin{figure}
\centerline{\psfig{figure=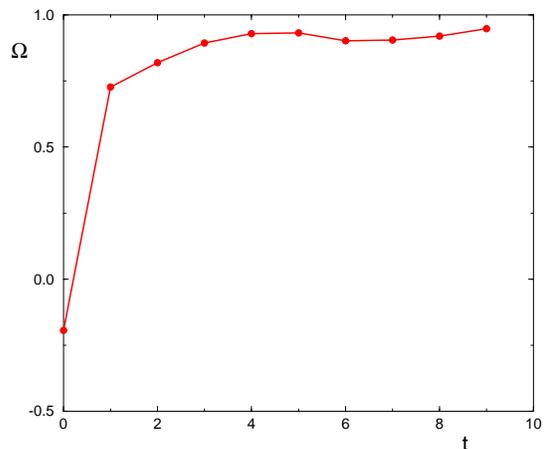,height=7.0cm,angle=270}}
\caption{Convergence of the scalar product 
$ \Omega = {\bf w}_t \cdot {\bf w}_{t+1}$
in the learning process, as a function of the epoch $t$.}
\label{fig:dot-all}
\end{figure}

Another aspect of the solutions derived by learning is their convergence.
We calculated, after every epoch $t$, 
the {\it average} solution (i.e. the center
of mass of version space), ${\bf w}_t$. 
The overlap $ \Omega = {\bf w}_t \cdot {\bf w}_{t+1}$
of such average solutions,
measured after two  successive epochs $t$ and $t+1$, 
increases with $t$ to a value very
close to 1 (see Fig. \ref{fig:dot-all}). This indicates 
that the vector ${\bf w}_t$ converges for large $t$s 
to some particular direction.

\section{Conclusions}

One of the simplest and more widely used forms
for the energy of a protein is the contact approximation,
(see Eq.(\ref{eq:contact0})).
Although such approximations have contributed a lot to our
understanding of the general properties of the folding transition,
it is far from clear which are its intrinsic limitations
if the task is to actually predict
native conformations from protein sequences.

To give a clear-cut solution to this problem,
in this work we have posed a remarkably simple question:
Is it possible to optimize a set of contact energy parameters
such that the native contact map of a {\it single} protein has
the lowest energy among all the possible alternative contact maps?

We have reached the conclusion that
the derivation of an such a set is
at the edge of learnability, even in the case of one protein only.
For any practical purpose, the basic requirement that 
the native state of a protein should be the minimum of some
effective free energy, cannot be fulfilled if a contact approximation
is used. 

We stress the obvious fact that
learning a set of contact energy parameters for more than one protein
is necessarily more difficult, and would not change our conclusion.

Our conclusion is supported by substantial evidence:
\begin{enumerate}
\item
If only two species of amino acids are used, the unlearnability
of the task is unambiguously shown.
\item
When 15 species of amino acids are considered, 
the time $\tau$ needed to learn the set of energy parameters 
that stabilize crambin increases dramatically
with the number $P$ of alternative contact maps that are considered.
Such increase of $\tau$ is compatible with a divergence at a finite $P$.
\item
The generalization error $\varepsilon_g$ does not asymptote to zero,
rather it fluctuates around a finite, although small, value.
\item
The distance from the native state of contact maps that are
found with energy lower than the native one
does not decrease as the optimization is carried on.
\item
The allowed region in energy parameter space shrinks to zero
along roughly a half of the total number of
directions. Thus, a further optimization
of parameters along these directions is ruled out.
\end{enumerate}

Even within a contact energy framework, more accurate and possibly
more successful approximations are possible.
For example, an all-atom based definition of contact 
instead of one based on the $C_{\alpha}$ only, could be expected to improve
the quality of the prediction.
We regard, however, the results presented here
as a first step towards a systematic improvement
of the approximation of the energy function to be used in folding predictions.
In planned future work, the simple form of the energy used here
will be supplemented with the inclusion of
additional energy terms,
such as hydrophobic (solvation), hydrogen bond or multi-body interactions.

A different question, which is not in the scope of the present work,
is how does a set of contact energy
parameters derived by perceptron learning compare with other existing sets.
We have addressed this problem by learning together 153 different proteins,
and considering alternative conformations generated by gapless threading
\cite{nvd97}.

This last issue is connected with the possibility to perform
predictions that do not rely completely on energy minimization alone.
For example, the condition that
the native state should be the absolute minimum of the energy function
can be relaxed.
In such ``weak'' prediction, 
a short list of candidates is identified and used as starting point
for a successive selection.

This research was supported by grants from the Minerva Foundation,
the Germany-Israel Science Foundation (GIF) and by a grant from the
Israeli Ministry of Science. 
We thank Ido Kanter for most helpful discussions.


\end{document}